\documentclass[prl,twocolumn,showpacs,floatfix,amsfonts]{revtex4}
\usepackage{graphicx,graphics,color,epsfig}
\usepackage{bm}
\usepackage{amsmath}
\usepackage{amssymb}

\newcommand{\bk}{{\bf k}}

\newcommand{\bsig}{\mbox{\boldmath{$\sigma$}}}

\begin{document}
\preprint{}
\title{ESR-STM of a single precessing spin: Detection of Exchange Based Spin
Noise}

\author{A. V. Balatsky$^1$, Yishay Manassen$^2$ and
Ran Salem$^2$ } \affiliation{$^1$Theoretical Division, Los Alamos
National Laboratory, Los Alamos, New Mexico 87545, $^2$ Department
of Physics and the Ilse Katz Center for Nanometer Scale Science
and Technology, Ben Gurion University, Beer Sheva, 84105, Israel}

\date{April 10, 2002}

\begin{abstract}
{ESR-STM is an emerging technique which is capable of detecting
the precession of a single spin. We discuss the mechanism  of
ESR-STM based on a direct exchange coupling between the tunneling
electrons and the local precessing spin $\bf S $. We claim that
since the number of tunneling electrons in a single precessing
period is small ($\sim 20$) one may expect a net temporary
polarization within this period that  will couple via exchange
interaction to the localized spin. This coupling will randomly
modulate the tunneling barrier and  create a dispersion in the
tunneling current which is a product of a Larmor frequency
component due to the precession of the single spin and the
dispersion of the spin of the tunneling electrons.   This noise
component is spread over the whole frequency range for random
white noise spin polarization of  electrons. In opposite case the
power spectrum of the spins of the tunneling electrons has a peak
at zero frequency an elevated noise in the current at $\omega_L$
will appear. We discuss the possible source of this spin
polarization. We find that for relevant values of parameters
signal to noise ratio in the spectral characteristic is 2-4 and is
comparable to the reported signal to noise ratio \cite{1,2}. The
magnitude of the current fluctuation is a relatively weak
increaing function of the DC current and the magnetic field. The
linewidth produced by the back action effect of tunneling
electrons on the precessing spin is also discussed.

 }
\end{abstract}

\maketitle

There is a growing realization that the technique of ESR-STM  is
capable of detecting the precession of a single surface spin by
modulating the tunneling current at the Larmor frequency. This
technique was successful in measuring Larmor frequency modulations
in defects in semiconductor surfaces \cite{1} and in paramagnetic
molecules \cite{2}. The increasing interest in this technique is
due to the possibility to detect and manipulate a single spin
\cite{3}.

 The alternative technique that allows one to detect single spin is the
optically detected magnetic resonance (ODMR) spectroscopy in a
single molecule \cite{4}. In comparison, ESR-STM has the unique
ability to correlate the spectroscopic information with the
spatial information, detected at the atomic level. It also  allows
one  to manipulate the position of the spin centers at the atomic
level \cite{5}.

 There has been
several proposals for the mechanism of detection. One is a
polarization of the mobile carriers through spin orbit coupling,
and modulation of the LDOS as a result of the precession \cite{6}.
Another one is  the interference between two resonant tunneling
components through the magnetic field splitted Zeeman levels
\cite{7}. Both of these mechanisms rely on a spin orbit coupling
to couple a local spin ${\bf S}$ to the conduction electrons and
have assumed {\em no spin polarization of tunneling electrons}.
Recently however, Durkan and Welland
 \cite{2} observed a
strong signal in a system with a substantially smaller spin orbit
coupling
 than what was assumed
in the calculations \cite{6},\cite{7}.
 Motivated by these experiments we addressed a
question: what is the
 role of the {\em direct} exchange interaction between
the localized
 spin and the tunneling electrons. Exchange interaction
has a
 tremendous influence on the physics of
conducting substances when magnetic impurities are present
\cite{8} and it is  natural to ask here: Does exchange interaction
play a role in ESR-STM also?

 We find that
 a direct
Heisenberg exchange interaction between the localized spin and the
conduction electrons is capable of producing the modulation of the
tunneling current. The qualitative difference compared with the
previous models is that we consider temporal fluctuations of the
spin polarization of the electrons that are tunneling between the
tip and the surface. Spin orbit interaction is irrelevant for this
consideration.  We argue in this paper, that although the spin
polarization of the tunneling electrons is zero in the long time
limit, it is not zero on the scale of the period of the
precession, typically $1/\omega_L \sim 2 ns$. On this time scale
there are very few electrons that pass near by the localized spin.
There exists a temporary spin polarization of the tunneling
electrons, which may interact, through exchange interaction with
the localized spin center.

\begin{figure}
\epsfig{figure=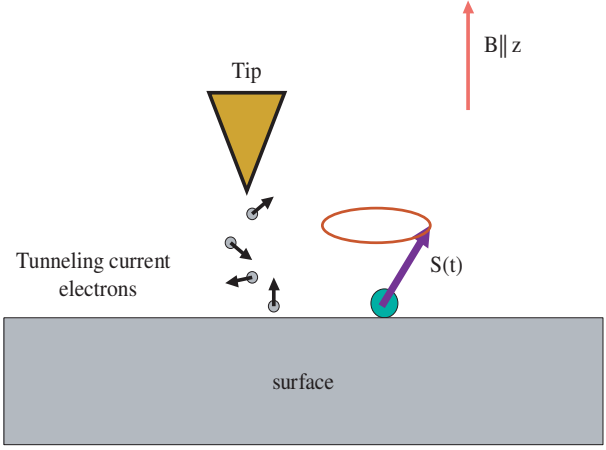,width=8 cm}
\caption[*]{  Schematics of the ESR-STM experiment is shown. The
fluctuations in the spin polarization of the tunneling electrons
at the time scale of the precession $T$ will be nonzero and will
scale as $\frac{1}{\overline{N}}$. $\overline{N} = 1/e I_0 T$ is
the average number of electrons tunneling between  tip and the
surface during one precession cycle. Once the tip is positioned
close to the localized spin, the exchange interaction between the
localized spin and the tunneling electrons will modulate the
tunneling current. The conditions in which this random modulation
will create a $\omega_L$ peak are discussed in the text.}
\label{FIG:STM}
\end{figure}

 It is important to point out that ESR-STM technique
performs a {\em noise
 spectroscopy}. We do not drive the single spin with
an external coherent rf field,  and  we are basically detecting an
incoherent phenomenon (we avoid here the question of the meaning
of this concept on a single particle level). There have been
several demonstrations in the past of detecting magnetic resonance
with noise spectroscopy \cite{9}. We argue that   it is  possible
to get a noise related signal from an exchange interaction between
the tunneling electrons and the localized surface spin center.

The overlap of the electron wave function in the tip and surface,
separated by a distance $d$ is exponentially small and is given by
a {\em spin dependent} tunneling matrix element:
\begin{eqnarray}\label{eq:G}
  \hat{\Gamma} = \Gamma_0 \exp[-\sqrt{\frac{\Phi -
J{\bf S}(t)
  {\hat{\bsig}}}{\Phi_0}}]
\end{eqnarray}
where we consider the spin ${\bf S}(t)$ in the magnetic field
$B||z$, precessing with the Larmor frequency $\omega_L  = g \mu_B
B$,  $\hat{\Gamma}$ is understood as a matrix in spin indexes,
$\Phi$ is the tunneling barrier height, $\Phi$ is typically few eV
and we assume $\Phi = 4 eV$, and $\Phi_0 = \frac{\hbar^2}{8 m
d^2}$ is the energy related to the distance between tip and
surface $d$ \cite{Stroscio}. The exchange term in the exponent is
small compared to the barrier height and we can expand the
exponent in $JS$. Explicitly $\hat{\Gamma}$ can be written as
\begin{eqnarray}\label{eq:G2}
\hat{\Gamma} = \Gamma_0 \exp(-(\Phi/\Phi_0)^{1/2})[
cosh[\frac{JS}{2\Phi} \sqrt{\frac{\Phi}{\Phi_0}}]
\nonumber\\
+\hat{\bsig} {\bf n(t)} sinh[\frac{JS}{2\Phi}
\sqrt{\frac{\Phi}{\Phi_0}}]]
\end{eqnarray}
where $\Gamma_0$ describes spin independent tunneling in the
absence of $J$. Note that the dynamics of the spin is now absorbed
in the time dependence of the unit vector ${\bf n}(t): {\bf S} =
{\bf n} S$. Let us now give a simple qualitative description of
the effect we address here. Leaving aside the constants we see
that the tunneling conductance  has a part that depends on the
localized spin
\begin{equation}\label{eq:I1}
  \delta I(t) \sim {\bf n}(t) \bsig (t)
\end{equation}
in a scalar product ${\bf n}(t) \bsig (t) = n^z(t) \sigma^z(t)
 + n^x(t) \sigma^x(t) + n^y(t) \sigma^y(t)$ only a
transverse part, that depends on the $x,y$ components of the
localized spin and the spin of the tunneling electrons, will
describe precession in a magnetic field ($B||z$ is assumed).  We
will focus on the transverse terms  below. To make the argument as
simple as possible we will assume at the moment that the spin
${\bf S}(t)$ is a simple periodic function of time $n_x(t) =
n_{\perp}\cos(\omega_L t), n_y(t)= n_{\perp} \sin(\omega_L t)$
with the period $T = 2\pi/\omega_L$. It is convenient to introduce
a time average of the current over $T$  $\Delta I = 1/N \sum_{i=
1}^N \delta I(t_i)$, where the sum over i=1 to N is over the
number of electrons that will tunnel between tip and the surface
in time $T$, with an average
 $ \overline{N} = I_0 T$,
which is dependent on the dc current in the system $I_0$.
\begin{equation}\label{eq:I2}
  \Delta I = \frac{1}{N} \sum_{i= 1}^N
\sigma^x(t_i)n^x(t_i) + (
  x\rightarrow y)
\end{equation}
This term represents the fluctuations of tunneling current due to
the interaction with the single precessing spin. Then the
dispersion of the current, that depends on the precessing
components is given by the dispersion of the quantity $\sum_{i,j=
1}^N n_x(t_i) n_x(t_j) (\sigma^x(t_i)\sigma^x(t_j))$. If the spin
wave functions of the tunneling electrons are not correlated
between different tunneling events we find
\begin{equation}\label{eq:disp}
(\sum_{i= 1}^N \sigma^x(t_i)n_x(t_i))^2  + (x \rightarrow y) \sim
 \overline{N}
\end{equation}
Therefore the dispersion of the current due to the exchange
interaction between the localized precessing spin and the spin of
the tunneling electrons is:
\begin{equation}\label{II}
\frac{\overline{\langle \Delta I^2 \rangle}}{I^2_0}\sim \langle
(n^{x})^2 \rangle \frac{\overline{N}}{\overline{N}^2} + ( x
\rightarrow y) \sim \frac{1}{\overline{N}}
\end{equation}
Where  the result is normalized to the dc current magnitude. We
find that the magnitude of the fluctuations $(\overline{\langle
\Delta I^2 \rangle})^{1/2}$ is on the scale of few percent of the
dc current for experimentally relevant values of parameters, see
Eq.(\ref{eq:varI}). If the spins of the tunneling electrons are
totally uncorrelated, this noise component will be smeared over
the whole frequency range. However, as we show below,  if some
tunneling electron spin polarization exists, a strong noise peak
will appear at $\omega_L$.

We argue that this simple mechanism is in agreement with several
experimental observations, such as the intensity of the signal and
the signal's linewidth. From Eq.(\ref{II}) we can immediately
conclude that the mean square fluctuation of the spin dependent
current is {\em a  weak increasing function of  both magnetic
field and dc current} with power $\frac{1}{2}$
\begin{equation}\label{eq:IB}
(\overline{\langle \Delta I^2 \rangle})^{1/2} \sim (I_0 B)^{1/2}
\end{equation}

We will now give a derivation of the results. Consider the set up
that is used in ESR STM, Fig.1. Since the tip is very close to the
magnetic site, we assume that the Heisenberg exchange coupling
between conduction electrons that tunnel across  the barrier and
the localized spin ${\bf S} = {\bf n} S$, is typically on the
order of a fraction of $eV$ .
 Hence the effective barrier, seen by
tunneling electron, will depend on the spin of the conduction
electron.

Let us first discuss relevant time scales of the problem.  For
$I_0 = e/\tau_e =  1 nA$ current electron tunneling rate is
$\frac{1}{\tau_e} \sim 10^{10} Hz$. The electron precession
frequency at field $B \sim 200 Gauss$ is about $ \omega_L/2 \pi =
500 MHz, T = 2 \times 10^{-9} sec$. Per single precession cycle
there are about $\overline{N} = 20$ electrons that tunnel between
the tip and the surface. As we indicated above, the fluctuation of
the electron spin is appreciable $\sim (\overline{N})^{1/2} \sim 4
$ for such a small number of electrons.

a) {\em Spin dependent tunneling}. We model the effect of
Heisenberg interaction as a spin dependent tunneling barrier. For
practical purposes we can assume that the precessing localized
spin ${\bf S}(t)$ is slow compared with the typical  tunneling
time of electron.

 The Hamiltonian we consider describes a spin
dependent tunnneling matrix element between the tip (L electrode)
and
 the surface (R electrode)
\begin{equation}\label{eq:Ham}
   H = \sum_{\bk,\alpha} \epsilon(\bk) c^{\dag}_{L
\alpha}(\bk) c_{L
  \alpha}(\bk) + (L \rightarrow R) + \sum_{\bk,\bk'}
c^{\dag}_{L \alpha}(\bk)\Gamma_{\alpha \beta}
  c_{R \beta}(\bk')
\end{equation}
 We assume  that the magnetic field is  along z axis:
$B||z$. The tunneling current operator will contain the spin
independent
 part that we omit
hereafter and the spin dependent part:
\begin{equation}\label{eq:J1}
 \delta \hat{I}(t) = \Gamma_1 {\bf n}(t) \bsig,
\end{equation}
where $ \Gamma_1 = \gamma_0
\sinh[\frac{JS}{2\Phi}(\Phi/\Phi_0)^{1/2}]$. We introduced a
  renormalized $\gamma_0 =
\Gamma_0\exp(-(\Phi/\Phi_0)^{1/2}$ that determines the
  dc current at a given bias $V$: $I_0 =\gamma_0 V$.
The current-current
 correlator, normalized to dc current  is then:
\begin{eqnarray}\label{eq:J2}
  \frac{\overline{\langle \delta \hat{I}(t) \delta
\hat{I}(t')\rangle}}{I^2_0} =
(\sinh[\frac{JS}{2\Phi}(\Phi/\Phi_0)^{1/2}])^2
\nonumber \\
\sum_{i,j = x,y,z}
  \langle n^i(t)n^j(t')\rangle \overline{\sigma^i(t)
\sigma^j(t')}
\end{eqnarray}
We explicitly separate the averaging over the dynamics of the
localized spin $\langle AB \rangle$ and the averaging over the
ensemble of the tunnelling electrons $\overline{AB}$. For the spin
dynamics we use $\langle n^x(t)n^x(t')\rangle \sim
\cos(\omega_L(t-t'))\exp(- \gamma |t-t'|)$ and similar for $y$
component. For the averaged over time $T$ current-current
correlator we will have result, similar to Eq.(\ref{eq:J2}) with
$\delta I \rightarrow \Delta I$  (see definition in
Eq.(\ref{eq:I2}) and above). This  brings an additional factor of
$\frac{1}{\overline{N}}$.

 To estimate the magnitude of the current fluctuations
due to the coupling to the localized spin we will take $J \sim 0.1
eV $.
 This is typical for an
exchange interaction in semiconductors and metals \cite{exchange}.
The
 barrier height $\Phi \simeq
4eV$ , spin $ S = 1/2$. To estimate $ \Phi_0 = \frac{\hbar^2}{8 m
d^2}$ we
 assume typical
tunneling distance $ d = 4 \AA $. This yields $\Phi_0 \simeq 0.1
eV$. For these parameters we find
\begin{equation}\label{eq:varI}
\frac{(\overline{\langle \Delta I^2 \rangle})^{1/2}}{I_0} \simeq
\frac{2}{\overline{\sqrt{N}}}
\sinh[\frac{JS}{2\Phi}(\Phi/\Phi_0)^{1/2}] \simeq 0.01
\end{equation}
 $\Gamma_1 = 0.02  \gamma_0$. The magnitude of the
fluctuation is in the $ 10 pA$ range for a tunneling current of
$I_0 = 1nA$ and is within the observed range \cite{1,2}. This is a
magnitude of  the fluctuating  current in  time domain due to the
exchange interaction between the precessing single spin and the
tunneling electrons. This current fluctuation will give a peak at
the Larmor frequency once there exist some spin polarization in
the tunneling current on a time scale of the relaxation time of
the single spin. For finite spin polarization, the size of the
noise component will be larger also. Thus the value of $10 pA$
represents a minimal intensity. Actual signals will increase with
the degree of spin polarization of the tunneling electrons.

b) {\em Back action effect of the tunneling current on the spin}.
One can use the tunneling Hamiltonian Eq.(\ref{eq:Ham}) to
estimate the decay rate of the localized spin state due to
interaction $\Gamma_1$. To second order this calculation is
equivalent  to the Fermi golden rule calculation and we have $
\frac{1}{\tau_s} = \pi \Gamma_1^2 N_L N_R eV$. Similarly, the DC
tunneling current $I_0$ is given by the tunneling rate of
conduction electrons $ \frac{1}{\tau_e} = \pi \gamma_0^2 N_L N_R
eV$, where $N_{L,R}$ is the density of states at the Fermi level
of the tip and surface respectively \cite{Korotkov}. One finds by
combining these two equations:
\begin{equation}\label{eq:taus}
\frac{1}{\tau_s} = \frac{1}{\tau_e}\frac{\Gamma_1^2}{\gamma_0^2}
\simeq
  4 \times 10
^{-4} \frac{1}{\tau_e}
\end{equation}
This result has a simple interpretation: The electron tunneling
rate $\frac{1}{\tau_e} \sim 10^{10} Hz$ gives the attempt rate for
the tunneling electrons. The probability to flip the localized
spins is proportional to $\Gamma_1^2$, which gives
Eq.(\ref{eq:taus}) for the linewidth. We estimate $
\frac{1}{\tau_s} \simeq 4 \times 10^{6} Hz $. This estimate is
within an order of magnitude of the reported linewidth \cite{1,2}.
Given the uncertainty in the parameters used we believe this is a
reasonable result; for example if we take $J = 0.05 eV$ we will
find $\frac{(\overline{\langle \Delta I^2 \rangle})^{1/2}}{I_0}
\sim 10^{-2}$ and linewidth will change by factor of 4 $
\frac{1}{\tau_s} \simeq  10^{6} Hz $. Linewidth will increase with
the increased spin polarization of tunneling electrons.  Future
experiments will help to clarify the linewidth dependence on $J,
B$ and other parameters.

In the above discussion,  we have assumed that the dynamics of the
local spin is controlled by the magnetic field only and no
decoherence mechanism, except back-action is included. In practice
there are other sources of dephasing of a precessing spin that
will add to the backaction effect of tunneling electrons and
details will depend on the specific material. In this context we
point out that the ESR linewidths are quite narrow  for magnetic
centers in semiconductors and insulators even at room
temperatures, typically few MHz, \cite{linewidth}. In the case of
a single spin linewidth will be narrower as the inhomogeneous
broadening is not an issue in this case.

For any source of decoherence, be it back action scattering or the
interaction with environment, the localized spin will be scattered
from the ground state and produce mixed states with nonzero
$<S_x>, <S_y>$, required to have precessing spin. No phase
coherence between different precessing spins is required as we are
looking at the single site.

c)  {\em Spectral density of the current}. The Fourier transform
of the current-current correlator will give a power spectrum of
the current fluctuation, Eq.(\ref{eq:J2}):
\begin{eqnarray}\label{eq:Ipower1}
 \frac{\overline{\langle I^2_{\omega} \rangle}}{I^2_0} =
  (\sinh[\frac{JS}{2\Phi}(\Phi/\Phi_0)^{1/2}])^2
\nonumber \\
\sum_{i = x,y,z} \int \frac{d\omega_1}{2\pi} \langle
(n^i)^2_{\omega - \omega_1}\rangle
\overline{(\sigma^i)^2_{\omega_1}}
\end{eqnarray}
where $\langle (n^i)^2_{\omega}\rangle \simeq
\frac{\gamma}{(\omega - \omega_L)^2 + \gamma^2} $ is the power
spectrum of ${\bf n}(t)$ fluctuations and
$\overline{(\sigma^i)^2_{\omega}} \simeq
\frac{\gamma_{\sigma}}{(\omega)^2 + \gamma_{\sigma}^2}$ is the
power spectrum of $\sigma^i(t)$ which we approximate as a
Lorenzian at zero frequency with the width given by the maximum
$\gamma_m = max(\gamma, \gamma_{\sigma})$. We get for a spectral
power density
\begin{eqnarray}\label{eq:Ipower2}
 \overline{\langle I^2_{\omega} \rangle} \simeq I^2_0
  (\sinh[\frac{JS}{2\Phi}(\Phi/\Phi_0)^{1/2}])^2
\nonumber \\  \frac{\gamma_m}{(\omega - \omega_L)^2 + \gamma_m^2}
\end{eqnarray}
Hereafter we omit subscript $m$ in $\gamma_m$ for simplicity. We
assume that $\gamma_{\sigma} \leq \gamma$ and $\gamma_m \simeq
\gamma$. It is useful to relate this spectral density to the shot
noise power spectrum $ \langle I^2_{shot}(\omega) \rangle = 2 e
I_0 \Delta \omega $. We have
\begin{equation}\label{eq:Ipower3}
  \frac{\overline{\langle I^2_{\omega} \rangle}}{\langle I^2_{shot}(\omega) \rangle} =
  (\sinh[\frac{JS}{2\Phi}(\Phi/\Phi_0)^{1/2}])^2 \frac{ 1/\tau_e
  \gamma}{(\omega - \omega_L)^2 + \gamma^2}
\end{equation}
At Larmor frequency we find that signal to noise ratio is
\begin{equation}\label{eq:Ipower4}
\frac{\overline{\langle I^2_{\omega} \rangle}}{\langle
I^2_{shot}(\omega) \rangle} =
(\sinh[\frac{JS}{2\Phi}(\Phi/\Phi_0)^{1/2}])^2 \frac{ 1}{\tau_e
\gamma} \simeq 2-4
\end{equation}
and we see that signal is large and certainly detectable and is
close to what has been observed experimentally. We used $\gamma
\sim 1/\tau_s = 1 Mhz$ for a linewidth, $1/\tau_e = 10^{10} Hz$
and $ \sinh[\frac{JS}{2\Phi}(\Phi/\Phi_0)^{1/2}] = 0.02$ for our
values of parameters. We also point out that the above analysis
could be equally applied to other configurations, say current in
nanostructures with no STM tunneling current.

In order to get a well defined signal at $\omega_L$ a $transverse$
spin polarization must exist in the tunneling current
(longitudinal components can not interact with the time dependent
components of the single spin). There are several possibilities in
which such a polarization might be created: The first obvious
possibility is due to the absorption of a paramagnetic atom or
cluster on the tip. The magnetic moment at the edge of the tip
will feel a strong magnetic anisotropy which will tend to force it
in the direction of the easy axis (not necessarily in the z
direction.) Such a paramagnetic tip can be a source of spin
polarized tunneling electrons in the transverse direction. One may
think of such a polarization even in the absence of a paramagnetic
tip. The time independent component of the single spin (which is
in the z-direction) may introduce a transverse polarization in
points where the spin polarization changes from parallel to
antiparallel. Such a phenomenon occurs for example with  a
dominant quadrupolar exchange interactions   where J goes through
zero as a function of distance \cite {slon}.
 Further work is
required to understand the mechanisms of polarizing the tunneling
electrons.

 As a direct outcome of this analysis we discuss the
 possible use of a paramagnetic tip. Tip of this
 sort can be
 prepared  by
 evaporating a
 thin magnetic  layer on it \cite {10}. Working with such tips
may enable a more well defined experiments where the spin
polarization of the tunneling electrons will be dependent on the
type of paramagnetic material deposited on the tips.

There are many possibilities to modify the tip material, from
working with a antiferromagnetic tip, to a superconducting (at low
temperatures) tip (for example made by Nb) to take advantage of
the Meissner effect, and to create a signal with stronger
intensities.

   In this paper we have shown that the temporal spin polarization of the
tunneling electrons can interact, through the Heisenberg exchange
interaction with the precessing spin. We have shown that such a
mechanism can create an elevated noise level at the Larmor
frequency with an intensity and linewidth which are comparable to
what is detected experimentally.

The potential scientific merit of this technique is very large.
Several milestones have to be achieved on different spin systems
to bring this technique to maturity: Detection of the  hyperfine
couplings; Observation of  ESR-STM signal from well defined
defects or atoms  on the surface, and observation of spin-spin
interactions from neighboring spins. After all these results are
shown it might be possible to prove that the ultimate goal,  a
single spin, could be indeed detected. It also would be very
interesting to observe the effect of an external excitation field
on the signal (excitation and saturation). Successful achievement
of these
 milestones will result in a very powerful technique with a broad range of
applications.

{\bf Acknowledgments}: This work was supported by the US
Department of Energy, by the German Israeli Foundation for
Research and Development (GIF) and by the Israel Science
Foundation (ISF). Y.M. wants to express his gratitude to the
theory group at Los Alamos for their kind hospitality.

\end{document}